\documentclass[conference]{IEEEtran}
\IEEEoverridecommandlockouts

\usepackage{amsmath,amssymb,amsfonts,amsthm}
\usepackage{tikz}
\usepackage[
  letterpaper,
  top=0.73in,
  bottom=1.12in,
  left=0.625in,
  right=0.625in
]{geometry}
\usetikzlibrary{arrows.meta,positioning}

\newtheorem{theorem}{Theorem}
\newtheorem{proposition}[theorem]{Proposition}
\newtheorem{lemma}[theorem]{Lemma}
\theoremstyle{remark}
\newtheorem{remark}[theorem]{Remark}

\newcommand{\CVaR}{\operatorname{CVaR}}
\newcommand{\KL}{D_{\mathrm{KL}}}
\newcommand{\kl}{\operatorname{kl}}

\newcommand{\cM}{\mathcal M}
\newcommand{\cP}{\mathcal P}
\newcommand{\bE}{\mathbb E}
\newcommand{\bP}{\mathbb P}
\newcommand{\1}{\mathbf 1}
\newcommand{\Alg}{\mathsf A}
\newcommand{\UCBVI}{\mathsf{UCBVI}}
\newcommand{\Reg}{\mathfrak R}
\newcommand{\Hist}{\mathcal H}

\begin{document}

\title{Worst-Case Bayesian CVaR Regret Bounds for Episodic Markov Decision Processes}

 \author{%
   \IEEEauthorblockN{
                     Raghav Bongole,
                     Amirreza Zamani,
                     Tobias J. Oechtering,
                      and Mikael Skoglund%
   \IEEEauthorblockA{
                    KTH Royal Institute of Technology, Stockholm, Sweden,
                     \{bongole, amizam, oech, skoglund\}@kth.se}
\thanks{This work is supported by the Knut and Alice Wallenberg Foundation.}
}}

\maketitle

\begin{abstract}
We study the worst-case, over priors, Bayesian conditional value-at-risk
(CVaR) of cumulative pseudo-regret in time-homogeneous finite-horizon
episodic Markov decision processes. A high-probability guarantee for
UCBVI-BF yields a prior-predictive upper bound. A lower bound follows by a three-state MDP and applying a reverse-KL inequality. For a constant number of states, the
bounds have the same polynomial dependence on horizon, actions, and
episodes, with a logarithmic gap in the risk level.
\end{abstract}

\begin{IEEEkeywords}
Bayesian regret, conditional value-at-risk, episodic reinforcement learning,
UCBVI, reverse KL
\end{IEEEkeywords}

\section{Introduction}

High-probability regret guarantees control the likelihood of rare, large losses, whereas Bayesian expected regret averages over all possible outcomes. Conditional value-at-risk (CVaR) complements these criteria by measuring the average regret in the worst \(1-\alpha\) fraction of the distribution. For a loss \(X\) and a risk level \(\alpha\in[0,1)\), it is defined by
\(
    \CVaR_\alpha(X)
    :=
    \inf_{\eta\in\mathbb R}
    \left\{
        \eta+\frac{1}{1-\alpha}
        \mathbb E\!\left[(X-\eta)_+\right]
    \right\},
\)
where \((x)_+:=\max\{x,0\}\). Thus, \(\CVaR_\alpha(X)\) accounts for both the probability and the magnitude of adverse outcomes. In particular, it can distinguish learning algorithms with similar expected regret but substantially different behavior on rare, high-regret events.

We study the smallest Bayesian CVaR regret achievable over \(K\) episodes under the least favorable prior on a class \(\cM\) of episodic MDPs. More formally, we consider
\begin{equation}
    \Reg_{K,\alpha}(\cM)
    :=
    \sup_{\mu\in\cP(\cM)}
    \inf_{\Alg}
    \CVaR_{\alpha}^{\mathbb P_{\mu,\Alg}}(R_K),
    \label{eq:minimax-value}
\end{equation}
Here, \(\cP(\cM)\) is the set of priors on \(\cM\), \(\Alg\) ranges over
admissible algorithms, and \(R_K\) is the cumulative pseudo-regret.
The law \(\mathbb P_{\mu,\Alg}\) includes the single draw \(M\sim\mu\),
the \(K\)-episode interaction, and all transition and algorithmic
randomness. T

For the upper bound, we use the algorithm called UCBVI-BF, introduced by Azar et al.~\cite{azar2017}. Although UCBVI-BF is a frequentist, prior-free algorithm, it remains a good candidate for the Bayesian optimization problem: for every prior, the best achievable Bayesian CVaR is no greater than that attained by any particular admissible algorithm. Moreover, the high-probability regret guarantee for UCBVI-BF holds uniformly over all MDPs in the class. We can therefore integrate its tail bound under an arbitrary prior and subsequently take the worst case over priors without changing the guarantee.

For the lower bound, we use the instantiation pattern for Bayesian CVaR
converses in \cite{bongole2026instantiating}. The three-state family is a bandit type construction simplified from \cite{dann2015sample}: only the first action is informative, and its
Bernoulli outcome determines all later rewards. 

\subsection{Contributions}
We give explicit upper and lower bounds for \eqref{eq:minimax-value}. The
upper bound converts the UCBVI-BF tail guarantee into a bound uniform over
priors. The lower bound transfers a bandit-like construction to episodic MDPs
and proves the resulting Bayesian CVaR bound by reverse KL. To our
knowledge, these are the first explicit bounds stated for the worst-case
prior-predictive CVaR of episodic-MDP pseudo-regret. The arguments establish a baseline and isolate what a
sharper, MDP-specific result must improve. In particular, for a constant number of states, the upper and lower
bounds on worst-case Bayesian CVaR regret match in their polynomial
dependence on \(H\), \(A\), and \(K\), and differ only by a factor of
\(\sqrt{\log(1/(1-\alpha))}\) in their dependence on the risk level.
Closing this logarithmic gap, and thereby obtaining a sharp
characterization, remains an open problem.

\subsection{Related work}

\paragraph{Minimax regret for episodic MDPs}
Azar et al. \cite{azar2017} introduced optimistic value-iteration algorithms
for tabular finite-horizon MDPs. UCBVI-BF has a model-uniform
high-probability bound with leading term
\(\widetilde O(\sqrt{HSAT})\), \(T=KH\), matching their minimax lower-bound
scale for this model class. Our argument uses its uniform tail statement. For the more general stage-dependent setting,
Jin et al. \cite{jin2018} gave an optimistic Q-learning bound and discussed
lower bounds built from the communicating-MDP construction of Jaksch et
al. \cite{jaksch2010}.

\paragraph{Information-theoretic bounds for interactive learning}
Chen et al. \cite{chen2024} developed interactive Fano, a reference-law
method unifying Le Cam, Assouad, classical Fano, and DEC-based converses for
adaptive data. Their fractional covering number measures how well one
exploration distribution covers near-optimal decisions; combined with the
DEC, it yields high-probability ExO+ bounds. Those results concern
frequentist minimax regret and quantiles, whereas our reverse-KL proof is a
direct prior-predictive event transfer for a finite hard family giving rise to worst case Bayesian CVaR bounds.

\paragraph{CVaR criteria}
Risk-sensitive bandit and RL papers often place CVaR on a different random
object. Tan and Weng \cite{tan2023} analyze the CVaR of bandit regret, while
Wang et al. \cite{wang2023} and Du et al. \cite{du2022} study
environment-wise or nested CVaR objectives for returns. By contrast,
\cite{generalizing2026} lower bounds CVaR under the joint prior-predictive
law, and \cite{bongole2026instantiating} instantiates this framework for
Gaussian estimation and bandits. We apply that viewpoint to episodic-MDP
pseudo-regret, take the worst case over priors, and strengthen the
risk-level dependence on $\alpha$ relative to
\cite{bongole2026instantiating}.

\section{Episodic MDP Setup}
\label{sec:setup}

\subsection{Finite-horizon interaction}

An episodic MDP is
\begin{equation}
 M=(\mathcal S,\mathcal A,H,P,r,\rho),
 \label{eq:mdp-tuple}
\end{equation}
where \(\mathcal S\) and \(\mathcal A\) are finite sets with
\(|\mathcal S|=S\) and \(|\mathcal A|=A\), \(H\ge2\) is the horizon,
\(P(\cdot\mid s,a)\) is the transition kernel,
\(r:\mathcal S\times\mathcal A\to[0,1]\) is a known deterministic reward
function, and \(\rho\) is the initial-state distribution. The same \(P\) and
\(r\) apply at every stage and the model is fixed across episodes. The class
\(\cM\) consists of MDPs satisfying these assumptions.

At the start of episode \(k\), the algorithm observes the history
\(\Hist_{k-1}\) and selects a possibly randomized, nonstationary policy
\(\pi_k=(\pi_{k,1},\ldots,\pi_{k,H})\). At stage \(h\),
\begin{equation}
 A_{k,h}\sim\pi_{k,h}(\cdot\mid S_{k,h}),\qquad
 S_{k,h+1}\sim P(\cdot\mid S_{k,h},A_{k,h}).
 \label{eq:trajectory-generation}
\end{equation}
The policy may depend on \(h\). The value functions are
\begin{align}
 V_{M,h}^{\pi}(s)
 &:=
 \bE_M^\pi\!\left[
   \sum_{\ell=h}^{H}r(S_\ell,A_\ell)
   \,\middle|\, S_h=s
 \right],
 \label{eq:policy-value}\\
 V_{M,h}^{*}(s)&:=\sup_{\pi}V_{M,h}^{\pi}(s).
 \label{eq:optimal-value}
\end{align}

\subsection{Pseudo-regret and prior-predictive law}

The cumulative pseudo-regret after \(K\) episodes is
\begin{equation}
 R_K(M)
 :=
 \sum_{k=1}^{K}
 \left[
 V_{M,1}^{*}(S_{k,1})
 -
 V_{M,1}^{\pi_k}(S_{k,1})
 \right].
 \label{eq:pseudo-regret}
\end{equation}
Since rewards belong to \([0,1]\),
\begin{equation}
 0\le R_K(M)\le KH.
 \label{eq:regret-bounded}
\end{equation}
Randomness in \(R_K\) comes from the sampled MDP, the algorithm, and the
observed trajectories. Let \(P_{\Alg}^{M}\) denote the transcript law
conditional on model \(M\). For a prior \(\mu\in\cP(\cM)\), define the
joint prior-predictive law \(\mathbb P_{\mu,\Alg}\) by
\begin{equation}
 \mathbb P_{\mu,\Alg}(dM,dh)
 :=\mu(dM)P_{\Alg}^{M}(dh).
 \label{eq:prior-predictive-law}
\end{equation}

\subsection{Conditional value-at-risk}

For an integrable loss \(X\) and \(\alpha\in[0,1)\), the
Rockafellar--Uryasev representation \cite{rockafellar2000} defines
upper-tail CVaR as
\begin{equation}
 \CVaR_\alpha(X)
 :=
 \inf_{t\in\mathbb R}
 \left\{
 t+\frac{1}{1-\alpha}\bE[(X-t)_+]
 \right\}.
 \label{eq:cvar-definition}
\end{equation}
Thus \eqref{eq:minimax-value} evaluates the tail of regret under the joint
law \eqref{eq:prior-predictive-law}; it does not mean that the learning
algorithm itself must be Bayesian.

\section{Main Results}
\label{sec:main}

\subsection{Upper bound from UCBVI-BF}

Let \(T:=KH\) and
\begin{equation}
 L_\delta:=\log\!\left(\frac{5HSAT}{\delta}\right).
 \label{eq:Ldelta}
\end{equation}
The following is the high-probability result used as a black box.

\begin{proposition}[UCBVI-BF]
\label{prop:ucbvi-hp}
Under the assumptions above and those of Theorem~2 in \cite{azar2017},
UCBVI-BF with confidence parameter \(\delta\in(0,1)\) satisfies, for every
\(M\in\cM\),
\begin{equation}
 P_{\UCBVI}^{M}(R_K>B_\delta)\le\delta,
 \label{eq:ucbvi-tail}
\end{equation}
where
\begin{align}
 B_\delta
 :={}&30HL_\delta\sqrt{SAK}
 +2500H^2S^2A L_\delta^2 \notag\\
 &+4H^{3/2}\sqrt{K L_\delta}.
 \label{eq:ucbvi-Bdelta}
\end{align}
\end{proposition}

\begin{theorem}[Worst-case Bayesian CVaR upper bound]
\label{thm:upper}
Fix \(\alpha\in[0,1)\), and define
\begin{equation}
 \delta_\alpha:=\frac{1-\alpha}{KH},
 \qquad
 L_\alpha
 :=\log\!\left(\frac{5HSAT\,KH}{1-\alpha}\right).
 \label{eq:upper-parameters}
\end{equation}
Then
\begin{align}
 \Reg_{K,\alpha}(\cM)
 \le{}&
 30HL_\alpha\sqrt{SAK}
 +2500H^2S^2A L_\alpha^2 \notag\\
 &+4H^{3/2}\sqrt{K L_\alpha}+1.
 \label{eq:upper-main}
\end{align}
\end{theorem}

\begin{remark}[Upper-bound method]
The model-uniform UCBVI-BF tail is first averaged under an arbitrary prior.
Lemma~\ref{lem:upper-conversion} then converts that prior-predictive tail
into a CVaR bound, and the choice
\(\delta=(1-\alpha)/(KH)\) controls the residual term. Taking the inner
infimum with UCBVI-BF and taking the supremum over priors proves the result.
When the first term leads, the rate is
\(\widetilde O(H\sqrt{SAK}\log(1/(1-\alpha)))\), where
\(\widetilde O\) hides logarithms independent of \(\alpha\).
\end{remark}

\subsection{Reverse-KL lower bound}

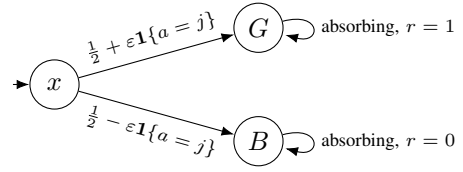
\begin{figure}[t]
\centering
\begin{tikzpicture}[
  >=Latex,
  state/.style={circle,draw,minimum size=6.5mm,inner sep=0pt},
  every edge quotes/.style={font=\scriptsize,align=center}]
 \node[state] (x) at (0,0) {$x$};
 \node[state] (g) at (2.7,0.75) {$G$};
 \node[state] (b) at (2.7,-0.75) {$B$};
 \draw[->] (-0.55,0) -- (x);
 \draw[->] (x) -- node[above,sloped,font=\scriptsize]
 {$\frac12+\varepsilon\mathbf 1\{a=j\}$} (g);
 \draw[->] (x) -- node[below,sloped,font=\scriptsize]
 {$\frac12-\varepsilon\mathbf 1\{a=j\}$} (b);
 \draw[->] (g) edge[loop right] node[right,font=\scriptsize]
 {absorbing, $r=1$} (g);
 \draw[->] (b) edge[loop right] node[right,font=\scriptsize]
 {absorbing, $r=0$} (b);
\end{tikzpicture}
\caption{Hard MDP \(M_j^\varepsilon\). Only the action chosen at \(x\)
is informative}
\label{fig:hard-family}
\end{figure}

For the lower bound, consider the three-state family in
Fig.~\ref{fig:hard-family}. The states and rewards are common to every MDP;
the index \(j\) only identifies the special first-stage action. Every
episode starts at \(x\), and the transition to \(G\) or \(B\) is the sole
random observation. The absorbing states repeat that Bernoulli outcome for
the remaining \(H-1\) stages, producing value gap
\((H-1)\varepsilon\).

\begin{theorem}[Bayesian CVaR lower bound]
\label{thm:lower}
Let \(H\ge2\), \(A\ge2\), \(K\ge1\), and \(\alpha\in(3/4,1)\). Every MDP
class containing the three-state hard family satisfies
\begin{align}
 \Reg_{K,\alpha}(\cM)
 &\ge
 \min\Biggl\{
 \frac{(H-1)K}{8}, \notag\\
 &\qquad
 \frac{H-1}{2\sqrt6}
 \sqrt{AK\log\frac{1}{4(1-\alpha)}}
 \Biggr\}.
 \label{eq:lower-main}
\end{align}
\end{theorem}

\begin{remark}[Lower-bound method]
The proof uses a uniform prior on the hard family and identifies an event
that has probability at least \(1/2\) under a reference MDP and incurs
regret of order \((H-1)\varepsilon K\). The adaptive KL chain rule bounds
the reverse-KL budget, after which reverse-KL event transfer and the
tail-to-CVaR lemma give a fixed-\(\varepsilon\) bound. Choosing
\(\varepsilon\) so that the transferred event fills the CVaR tail yields, \(\Omega(H\sqrt{AK\log(1/(1-\alpha))})\) for \(S=3\).
\end{remark}

\section{Discussion}

Theorem~\ref{thm:upper} is a worst-case Bayesian guarantee achieved by a
frequentist algorithm. Its validity depends on the model-uniform nature of
\eqref{eq:ucbvi-tail}, not on UCBVI-BF using a prior.

For \(S=3\), the leading upper and lower dependences are
\begin{align}
 \text{upper:}\quad&
 \widetilde O\!\left(
 H\sqrt{AK}\log\frac1{1-\alpha}\right),\notag\\
 \text{lower:}\quad&
 \Omega\!\left(
 H\sqrt{AK\log\frac1{1-\alpha}}\right).
 \label{eq:comparison}
\end{align}

Thus, for fixed \(S=3\) and fixed risk level, the bounds are asymptotically
close in \(H,A,K\): both have the polynomial scale \(H\sqrt{AK}\), up to
logarithmic and lower-order terms. However, a factor
\(\sqrt{\log(1/(1-\alpha))}\) remains between them. The converse captures
horizon dependence but not uncertainty across multiple states, and
therefore has no nontrivial \(S\)-dependence.

Two extensions are natural. Integrating the ExO high-probability bounds and
fractional-covering analysis of \cite{chen2024} would extend the upper bound
beyond tabular UCBVI; a risk-dependent ExO objective may improve the
\(\alpha\)-dependence. A communicating hard family in the spirit of
\cite{jaksch2010} may add state dependence to the converse, but requires
controlling policy-dependent state--action occupancies in the reverse-KL
argument.

\section{Proofs}
\subsection{Upper bound}
\label{app:upper}

We first prove the deterministic conversion used in the argument.

\begin{lemma}[One-level tail-to-CVaR conversion]
\label{lem:upper-conversion}
For every \(\alpha\in[0,1)\) and \(\delta\in(0,1)\), let
\(0\le X\le L_{\max}\) almost surely. If
\(\bP(X>B_\delta)\le\delta\), then
\begin{equation}
 \CVaR_\alpha(X)
 \le
 B_\delta+\frac{L_{\max}\delta}{1-\alpha}.
 \label{eq:upper-conversion}
\end{equation}
\end{lemma}

\begin{proof}
Choose \(t=B_\delta\) in \eqref{eq:cvar-definition}. Since
\((X-B_\delta)_+\le L_{\max}\1\{X>B_\delta\}\),
\begin{align}
 \CVaR_\alpha(X)
 &\le B_\delta+\frac{\bE[(X-B_\delta)_+]}{1-\alpha}\notag\\
 &\le B_\delta+\frac{L_{\max}\delta}{1-\alpha}.
 \label{eq:upper-conversion-step1}
\end{align}
\end{proof}

\begin{proof}[Proof of Theorem~\ref{thm:upper}]
Fix \(\alpha\in[0,1)\) and \(\delta\in(0,1)\).

\paragraph{Prior-predictive tail}
Proposition
\ref{prop:ucbvi-hp} states that, for every fixed \(M\in\cM\),
\begin{equation}
 P_{\UCBVI}^{M}(R_K>B_\delta)\le\delta,
 \label{eq:upper-proof-pointwise}
\end{equation}
with \(B_\delta\) given by \eqref{eq:ucbvi-Bdelta}. Crucially, the same
threshold \(B_\delta\) applies to every \(M\); it does not depend on the
prior or on the particular MDP. For any prior \(\mu\in\cP(\cM)\), the law
of total probability under
\eqref{eq:prior-predictive-law},
\begin{align}
 \mathbb P_{\mu,\UCBVI}(R_K>B_\delta)
 &=\int_{\cM}P_{\UCBVI}^{M}(R_K>B_\delta)\,\mu(dM)\notag\\
 &\le\int_{\cM}\delta\,\mu(dM)=\delta,
 \label{eq:upper-proof-mixture-tail}
\end{align}
because \(\mu\) is a probability measure. Thus the model-uniform
frequentist inequality is prior-predictive for every prior.

\paragraph{CVaR conversion and optimization}
Under the joint law \(\mathbb P_{\mu,\UCBVI}\), boundedness
\eqref{eq:regret-bounded} gives \(0\le R_K\le KH\). Apply Lemma
\ref{lem:upper-conversion} to \(X=R_K\), \(L_{\max}=KH\), and the tail bound
\eqref{eq:upper-proof-mixture-tail}. We obtain
\begin{equation}
 \CVaR_\alpha^{\mathbb P_{\mu,\UCBVI}}(R_K)
 \le
 B_\delta+\frac{KH\delta}{1-\alpha}.
 \label{eq:upper-proof-cvar-general-delta}
\end{equation}
Set \(\delta=\delta_\alpha=(1-\alpha)/(KH)\in(0,1)\). Then
\(KH\delta_\alpha/(1-\alpha)=1\) and
\(L_{\delta_\alpha}=L_\alpha\). Substitution into
\eqref{eq:upper-proof-cvar-general-delta} gives the right-hand side of
\eqref{eq:upper-main}, uniformly in \(\mu\). For each \(\mu\), UCBVI-BF is
an admissible algorithm, even though it does not use \(\mu\). Therefore
\begin{align}
 &\sup_\mu\inf_{\Alg}
 \CVaR_\alpha^{\mathbb P_{\mu,\Alg}}(R_K)\notag\\
 &\quad\le
 \sup_\mu
 \CVaR_\alpha^{\mathbb P_{\mu,\UCBVI}}(R_K)
 \le \text{right-hand side of \eqref{eq:upper-main}}.
 \label{eq:upper-proof-supremum}
\end{align}
This proves \eqref{eq:upper-main}.
\end{proof}

\subsection{Lower bound}
\label{app:lower}

\subsubsection{Hard MDP family}

Fix \(A\ge2\) and \(\varepsilon\in(0,1/4]\). The state space is
\begin{equation}
 \mathcal S=\{x,G,B\}.
 \label{eq:lower-states}
\end{equation}
The labels indicate the roles of the states: \(x\) is the common initial
state, \(G\) is good, and \(B\) is bad. Every episode starts from \(x\), so
\(\rho=\delta_x\). For each \(j\in[A]\), define \(M_j^\varepsilon\) by
\begin{align}
 \mathsf P_{j}^{\varepsilon}(G\mid x,a)
 &=
 \frac12+\varepsilon\1\{a=j\},
 \label{eq:lower-transition-G}\\
 \mathsf P_{j}^{\varepsilon}(B\mid x,a)
 &=
 \frac12-\varepsilon\1\{a=j\}.
 \label{eq:lower-transition-B}
\end{align}
For every \(a\in[A]\), the states \(G\) and \(B\) are absorbing:
\begin{equation}
 \mathsf P_j^\varepsilon(G\mid G,a)=1,
 \qquad
 \mathsf P_j^\varepsilon(B\mid B,a)=1.
 \label{eq:lower-absorbing}
\end{equation}
The reward function, common to the family, is
\begin{equation}
 r(x,a)=0,\qquad r(G,a)=1,\qquad r(B,a)=0.
 \label{eq:lower-rewards}
\end{equation}
The reference MDP \(M_0\) has the same rewards and absorbing transitions,
but
\begin{equation}
 \mathsf P_{0}(G\mid x,a)=\mathsf P_{0}(B\mid x,a)=\frac12
 \quad\text{for every }a\in[A].
 \label{eq:reference-mdp}
\end{equation}

Let \(\mathcal F_{k-1}:=\sigma(\Hist_{k-1})\) be the sigma-algebra generated
by the transcript before episode \(k\), and write \(A_k:=A_{k,1}\). Define
\begin{equation}
 p_{k,j}:=\Pr_{\Alg}(A_k=j\mid\mathcal F_{k-1}),
 \qquad W_j(K):=\sum_{k=1}^{K}p_{k,j}.
 \label{eq:soft-count}
\end{equation}
Only the first action appears because later actions cannot change an
absorbing state. The predictable soft count \(W_j(K)\) is the compensator
of \(N_j(K):=\sum_k\1\{A_k=j\}\): it sums conditional probabilities, not
realized indicators. For every history,
\begin{equation}
 \sum_{j=1}^{A}W_j(K)
 =
 \sum_{k=1}^{K}\sum_{j=1}^{A}p_{k,j}
 =K.
 \label{eq:soft-count-sum}
\end{equation}

\begin{lemma}[Regret identity]
\label{lem:regret-identity}
Under \(M_j^\varepsilon\), the cumulative pseudo-regret is
\begin{equation}
 R_{K,j}
 =(H-1)\varepsilon\bigl(K-W_j(K)\bigr).
 \label{eq:regret-identity}
\end{equation}
In particular, \(0\le R_{K,j}\le(H-1)\varepsilon K\).
\end{lemma}

\begin{proof}
If the first-stage action is \(j\), the probability of entering \(G\) is
\(1/2+\varepsilon\); every other action enters \(G\) with probability
\(1/2\). Since being in \(G\) yields one reward at each of the remaining
\(H-1\) stages, the value difference between action \(j\) and every action
\(a\ne j\) is
\begin{equation}
 g_\varepsilon
 =(H-1)\left[
 \left(\frac12+\varepsilon\right)-\frac12
 \right]
 =(H-1)\varepsilon.
 \label{eq:hard-gap}
\end{equation}
The policy value makes the conditioning explicit:
\begin{align}
 V_{M_j^\varepsilon,1}^{\pi_k}(x)
 &=(H-1)\left[
 p_{k,j}\left(\frac12+\varepsilon\right)
 +(1-p_{k,j})\frac12\right],\notag\\
 V_{M_j^\varepsilon,1}^{*}(x)
 &=(H-1)\left(\frac12+\varepsilon\right).
 \label{eq:hard-values}
\end{align}
The first line is the law of total expectation over whether \(A_k=j\), not
a mixture over MDPs. The episode-
\(k\) pseudo-regret is therefore
\begin{equation}
 g_\varepsilon(1-p_{k,j}).
 \label{eq:one-episode-hard-regret}
\end{equation}
Summing \eqref{eq:one-episode-hard-regret} over \(k=1,\ldots,K\) and using
\eqref{eq:soft-count} proves \eqref{eq:regret-identity}. The bounds follow
from \(0\le W_j(K)\le K\).
\end{proof}

For \(\varepsilon\in(0,1/4]\), let
\begin{equation}
 \mu_\varepsilon
 :=
 \frac1A\sum_{j=1}^{A}\delta_{M_j^\varepsilon}.
 \label{eq:hard-prior}
\end{equation}

\subsubsection{Auxiliary inequalities}

\begin{lemma}[Reverse-KL event transfer]
\label{lem:reverse-kl-event}
Let \(Q\) and \(P\) be probability measures, and let \(E\) be an event. If
\(Q(E)\ge1/2\), then
\begin{equation}
 P(E)
 \ge
 \frac14\exp\{-2\KL(Q\Vert P)\}.
 \label{eq:reverse-kl-event}
\end{equation}
\end{lemma}

\begin{proof}
Write \(q:=Q(E)\) and \(p:=P(E)\). If \(p=0\) while \(q>0\), then
\(\KL(Q\Vert P)=+\infty\) and \eqref{eq:reverse-kl-event} is trivial.
For \(p>0\), data processing under
\(\omega\mapsto\1\{\omega\in E\}\) and binary entropy
\(h(q)\le\log2\) give
\begin{align}
 \KL(Q\Vert P)
 &\ge\kl(q\Vert p)
 \ge q\log\frac1p-h(q)\notag\\
 &\ge\frac12\log\frac1p-\log2.
 \label{eq:p-log-bound-final}
\end{align}
Rearranging and exponentiating proves \eqref{eq:reverse-kl-event}.
\end{proof}

\begin{lemma}[Tail event to CVaR]
\label{lem:lower-event-cvar}
Let \(X\ge0\), let \(E\subseteq\{X\ge r\}\), and write \(p:=\bP(E)\). Then
\begin{equation}
 \CVaR_\alpha(X)
 \ge
 r\min\left\{1,\frac{p}{1-\alpha}\right\}.
 \label{eq:lower-event-cvar}
\end{equation}
\end{lemma}

\begin{proof}
Monotonicity gives
\(\CVaR_\alpha(X)\ge\CVaR_\alpha(r\1_E)\). For \(t\in[0,r]\), the
objective in \eqref{eq:cvar-definition} for \(Y=r\1_E\) is
\begin{align}
 t+\frac{p(r-t)}{1-\alpha}
 =
 \frac{pr}{1-\alpha}
 +t\left(1-\frac{p}{1-\alpha}\right).
 \label{eq:bernoulli-cvar-objective}.
\end{align}
The affine function is minimized at \(t=0\) if \(p\le1-\alpha\), and at
\(t=r\) otherwise. Values outside \([0,r]\) cannot improve either endpoint.
Therefore
\begin{equation}
 \CVaR_\alpha(r\1_E)
 =
 r\min\left\{1,\frac{p}{1-\alpha}\right\}.
 \label{eq:bernoulli-cvar-value}
\end{equation}
\end{proof}

\subsubsection{Proof of the lower bound}

\begin{proof}[Proof of Theorem~\ref{thm:lower}]
Fix \(\alpha\in(3/4,1)\), an arbitrary algorithm \(\Alg\), and an arbitrary
\(\varepsilon\in(0,1/4]\). Let \(Z\) be the complete transcript and write
\(Q^{0}:=P_{\Alg}^{M_0}\) and
\(P^{j}:=P_{\Alg}^{M_j^\varepsilon}\).
Superscripts indicate laws conditional on a fixed MDP. Define joint laws on
\((J,Z)\) by
\begin{align}
 \overline Q(J=j,Z\in D)&:=\frac1A Q^{0}(D),\notag\\
 \overline P(J=j,Z\in D)&:=\frac1A P^{j}(D).
 \label{eq:joint-laws}
\end{align}
Thus \(J\) is uniform and independent of \(Z\) under \(\overline Q\). Under
\(\overline P\), draw \(J\) uniformly, set \(M=M_J^\varepsilon\), and draw
\(Z\sim P^{J}\). Consequently, for every \(C\subseteq\cM\),
\begin{equation}
 \overline P(M\in C)
 =\frac1A\sum_{j=1}^{A}\1\{M_j^\varepsilon\in C\}
 =\mu_\varepsilon(C).
 \label{eq:mdp-marginal}
\end{equation}
More generally,
\begin{align}
 \overline P(M\in C,Z\in D)
 &=\frac1A\sum_{j=1}^{A}
 \1\{M_j^\varepsilon\in C\}P^{j}(D)\notag\\
 &=\int_C P_{\Alg}^{M}(D)\,\mu_\varepsilon(dM).
 \label{eq:pushforward-joint}
\end{align}
Thus the pushforward of \(\overline P\) to \((M,Z)\) is
\(\mathbb P_{\mu_\varepsilon,\Alg}\).

\paragraph*{Step 1: A frequent large-regret event}
Define \(E:=\{(j,z):W_j(K;z)\le K/2\}\).
For every transcript \(z\), \eqref{eq:soft-count-sum} implies that at most
one index can satisfy \(W_j(K)>K/2\). Therefore at least \(A-1\) indices
satisfy \(W_j(K)\le K/2\). Conditioning on \(Z\) under \(\overline Q\),
\begin{align}
 \overline Q(E)
 &=
 \int
 \frac1A\sum_{j=1}^{A}
 \1\!\left\{W_j(K;h)\le\frac K2\right\}
 Q^{0}(dz)\notag\\
 &\ge
 \int\frac{A-1}{A}\,Q^{0}(dz)
 =
 \frac{A-1}{A}
 \ge\frac12.
 \label{eq:reference-event-large}
\end{align}
Moreover, Lemma~\ref{lem:regret-identity} shows that on \(E\),
\begin{align}
 R_{K,j}
 &=(H-1)\varepsilon\bigl(K-W_j(K)\bigr)\notag\\
 &\ge
 \frac{(H-1)\varepsilon K}{2}
 =:r_\varepsilon.
 \label{eq:large-regret-threshold}
\end{align}
The boundedness part is
\( 0\le R_{K,j}\le(H-1)\varepsilon K.
\)

\paragraph*{Step 2: Reverse-KL budget}
Let \(Y_k:=\1\{S_{k,2}=G\}\) denote the transition outcome in episode \(k\),
and let
 \(\mathcal G_k:=\sigma(\Hist_{k-1},A_k)\).
Conditional on \(\mathcal G_k\), if \(A_k=a\), then
\begin{align}
 Q^{0}(Y_k=1\mid\mathcal G_k)&=\frac12,\notag\\
 P^{j}(Y_k=1\mid\mathcal G_k)
 &=\frac12+\varepsilon\1\{a=j\}.
 \label{eq:Y-conditionals}
\end{align}
Applying the KL chain rule sequentially to the transcript gives
\begin{align}
&\KL(Q^{0}\Vert P^{j})\\
&=\sum_{k=1}^{K}\bE_{Q^{0}}\!\left[
 \KL\!\left(Q^{0}(A_k\mid\mathcal F_{k-1})
 \Vert P^{j}(A_k\mid\mathcal F_{k-1})\right)\right.\notag\\[-2pt]
&\hspace{27mm}\left.+
 \KL\!\left(Q^{0}(Y_k\mid\mathcal G_k)
 \Vert P^{j}(Y_k\mid\mathcal G_k)\right)\right].
\label{eq:kl-chain-rule}
\end{align}
The action term is zero because, conditional on the same realized history,
the algorithm uses the same action kernel under both models. By
\eqref{eq:Y-conditionals}, the observation term is zero unless \(A_k=j\),
when it equals \(d(\varepsilon)\). Hence, by the tower property,
\begin{align}
 \bE_{Q^{0}}[\1\{A_k=j\}]
 &=
 \bE_{Q^{0}}[p_{k,j}],
 \label{eq:tower-action-count}\\
 \KL(Q^{0}\Vert P^{j})
 &=
 d(\varepsilon)\bE_{Q^{0}}[W_j(K)].
 \label{eq:individual-reverse-kl}
\end{align}

The Bernoulli divergence is
\(d(\varepsilon)=-\frac12\log(1-4\varepsilon^2)\). For
\(0<\varepsilon\le1/4\), the inequality
\(-\log(1-x)\le x/(1-x)\), with \(x=4\varepsilon^2\le1/4\), gives
\(d(\varepsilon)\le2\varepsilon^2/(1-4\varepsilon^2)
\le3\varepsilon^2\).

Since the marginals are the same uniform law, the conditional KL
chain rule for \((J,Z)\) gives
\begin{align}
 \KL(\overline Q\Vert\overline P)
 &=\frac1A\sum_{j=1}^{A}\KL(Q^{0}\Vert P^{j})\notag\\
 &=
 \frac{Kd(\varepsilon)}A
 \le
 \frac{3K\varepsilon^2}{A}.
 \label{eq:joint-reverse-kl-budget}
\end{align}
The factors \(1/A\) cancel inside the likelihood
ratio, while the outer expectation averages the conditional KL
divergences. The final equality uses \eqref{eq:soft-count-sum}.

\paragraph*{Step 3: Event transfer and CVaR}
Set \(p_\varepsilon:=\overline P(E)\). By
\eqref{eq:reference-event-large}, the premise of
Lemma~\ref{lem:reverse-kl-event} holds with \(Q=\overline Q\) and
\(P=\overline P\). Using \eqref{eq:joint-reverse-kl-budget} we get,
\begin{align}
 p_\varepsilon
 &\ge
 \frac14
 \exp\{-2\KL(\overline Q\Vert\overline P)\}
 \ge
 \frac14
 \exp\left(-\frac{6K\varepsilon^2}{A}\right).
 \label{eq:hard-event-probability}
\end{align}
By \eqref{eq:pushforward-joint}, the regret under \(\overline P\) has law
\(\mathbb P_{\mu_\varepsilon,\Alg}\). Since
\eqref{eq:large-regret-threshold} gives
\(E\subseteq\{R_K\ge r_\varepsilon\}\), Lemma
\ref{lem:lower-event-cvar} yields
\begin{align}
 &\CVaR_\alpha^{\mathbb P_{\mu_\varepsilon,\Alg}}(R_K)
\ge
 r_\varepsilon
 \min\left\{1,\frac{p_\varepsilon}{1-\alpha}\right\}\notag\\
 &\quad\ge
 \frac{(H-1)\varepsilon K}{2}
 \min\left\{
 1,
 \frac{
 \exp(-6K\varepsilon^2/A)
 }{4(1-\alpha)}
 \right\}.
 \label{eq:fixed-epsilon-lower}
\end{align}
This holds for every \(\varepsilon\in(0,1/4]\) and every algorithm.

\paragraph*{Step 4: Tail filling and worst-case value}
Put \(\beta:=1-\alpha\). In \eqref{eq:fixed-epsilon-lower}, the full regret
level \(r_\varepsilon\) is retained whenever
\(\frac14e^{-6K\varepsilon^2/A}\ge\beta\), equivalently
\(\varepsilon^2\le(A/(6K))\log(1/(4\beta))\). Since \(\alpha>3/4\), let
\(\Lambda_\alpha:=\log(1/(4(1-\alpha)))>0\) and choose
\begin{equation}
 \varepsilon_*
 :=
 \min\left\{
 \frac14,
 \sqrt{\frac{A\Lambda_\alpha}{6K}}
 \right\}.
 \label{eq:epsilon-star}
\end{equation}
Then \(6K\varepsilon_*^2/A\le\Lambda_\alpha\), so
\(e^{-6K\varepsilon_*^2/A}/(4(1-\alpha))\ge1\). Therefore the minimum in
\eqref{eq:fixed-epsilon-lower} equals 1, and
\begin{align}
 &\CVaR_\alpha^{\mathbb P_{\mu_{\varepsilon_*},\Alg}}(R_K)\ge
 \frac{(H-1)K}{2}
 \min\left\{
 \frac14,
 \sqrt{\frac{A\Lambda_\alpha}{6K}}
 \right\}\notag\\
 &\quad=
 \min\left\{
 \frac{(H-1)K}{8},
 \frac{H-1}{2\sqrt6}
 \sqrt{AK\Lambda_\alpha}
 \right\}.
 \label{eq:optimized-prior-lower}
\end{align}
The algorithm was arbitrary, so we may take its infimum. Since
\(\mu_{\varepsilon_*}\) is admissible in \eqref{eq:minimax-value}, its value
lower bounds the supremum over priors. Substituting the definition of
\(\Lambda_\alpha\) in \eqref{eq:optimized-prior-lower} proves
\eqref{eq:lower-main}.
\end{proof}

\section{Conclusion}

We established upper and lower bounds on worst-case Bayesian CVaR regret
for episodic MDPs. The upper bound converts the high-probability guarantee
of UCBVI-BF, while the converse applies reverse KL to a three-state MDP
family. For fixed \(S\), the bounds match polynomially in \(H,A,K\), but
differ by a factor of \(\sqrt{\log(1/(1-\alpha))}\) in their risk-level
dependence. The lower bound also has no nontrivial
\(S\)-dependence. Closing these gaps likely requires a multistage
hard family in which uncertainty and regret accumulate across states and
stages.

\newpage
\section*{AI Statement}
The main ideas of the paper were conceived by the authors without substantial aid of generative models.
OpenAI Codex was used to assist with organization, mathematical typesetting, language editing of the manuscript along with aiding with the algebraic derivations in Section V. Codex was also used to create the TikZ-code for figure. The authors are responsible for,
and have independently verified, all definitions, calculations, proofs, and
citations.

\IEEEtriggeratref{4}
\bibliographystyle{IEEEtran}
\bibliography{references}

\end{document}